# Simulating 3-D Stellar Hydrodynamics using PPM and PPB Multifluid Gas Dynamics on CPU and CPU+GPU Nodes

Paul R. Woodward[1], Pei-Hung Lin[2], Huaqing Mao[1], Robert Andrassy[3] and Falk Herwig[3]

1. Minnesota Institute for Astrophysics, University of Minnesota
2. Lawrence Livermore National Laboratory, Livermore, CA
3. University of Victoria, Victoria, B. C., Canada

paul@lcse.umn.edu

**Abstract**. The special computational challenges of simulating 3-D hydrodynamics in deep stellar interiors are discussed, and numerical algorithmic responses described. Results of recent simulations carried out at scale on the NSF's Blue Waters machine at the University of Illinois are presented, with a special focus on the computational challenges they address. Prospects for future work using GPU-accelerated nodes such as those on the DoE's new Summit machine at Oak Ridge National Laboratory are described, with a focus on numerical algorithmic accommodations that we believe will be necessary.

## 1. Introduction

Our team has been simulating brief events in stars that require a 3-D treatment. These events involve the action of convection in the transport of nuclear fuel in the stellar interior. Within a convection zone, the material rapidly becomes well-mixed, but at the upper and lower boundaries of a convection zone, mixing is much more difficult to describe. In a thin region of radii, the convective motions must cease, and buoyant material located just outside the convection zone resists becoming incorporated into the convection flow but can nevertheless be pulled into it in small concentrations. Attempts to model this process of convective boundary mixing (CBM) in 1-D stellar evolution codes have difficulty establishing a convincing validity in the absence of detailed 3-D simulations. Such simulations are needed to describe the multiple nonlinear processes that cause stably stratified gas to become entrained in a turbulent convection zone. We have attempted to fill this gap in our knowledge by performing such simulations – a snap shot from one of them is shown in Fig. 1 – and by generating the averages on spherical surfaces that can inform modeling efforts. This would be a detail in stellar evolution theory if it were not the case that entrainment of stably stratified gas can have dramatic effects on the behavior of the star under a number of special conditions. Cases where this process can be pivotal occur when the boundary of a convection zone approaches a source of especially combustible fuel [1-5] (or, in a mirror circumstance, when it approaches material that, if ingested, can dampen or even extinguish a nuclear flame [6,7]).

We have given special attention to ingestion of *hydrogen* fuel, because the unusually energetic combustibility of hydrogen can give rise to especially large energy release. Even when burning hydrogen in very small concentrations, the energy produced can rival the energy generated in shells where heavier elements such as helium are consumed. The reason for this is the very strong dependence of nuclear reaction rates on the temperature. Just above the helium burning shell, in the convection zone driven by this combustion, temperatures are very much higher than in the hydrogen-rich layers above the convection zone. The hydrogen is pulled downward by the convection until it reaches a level that is so hot that it burns rapidly enough to be consumed before descending still further. At these reaction rates, even small concentrations are sufficient to dramatically alter the convection flow and the overall luminosity of this region of the star.

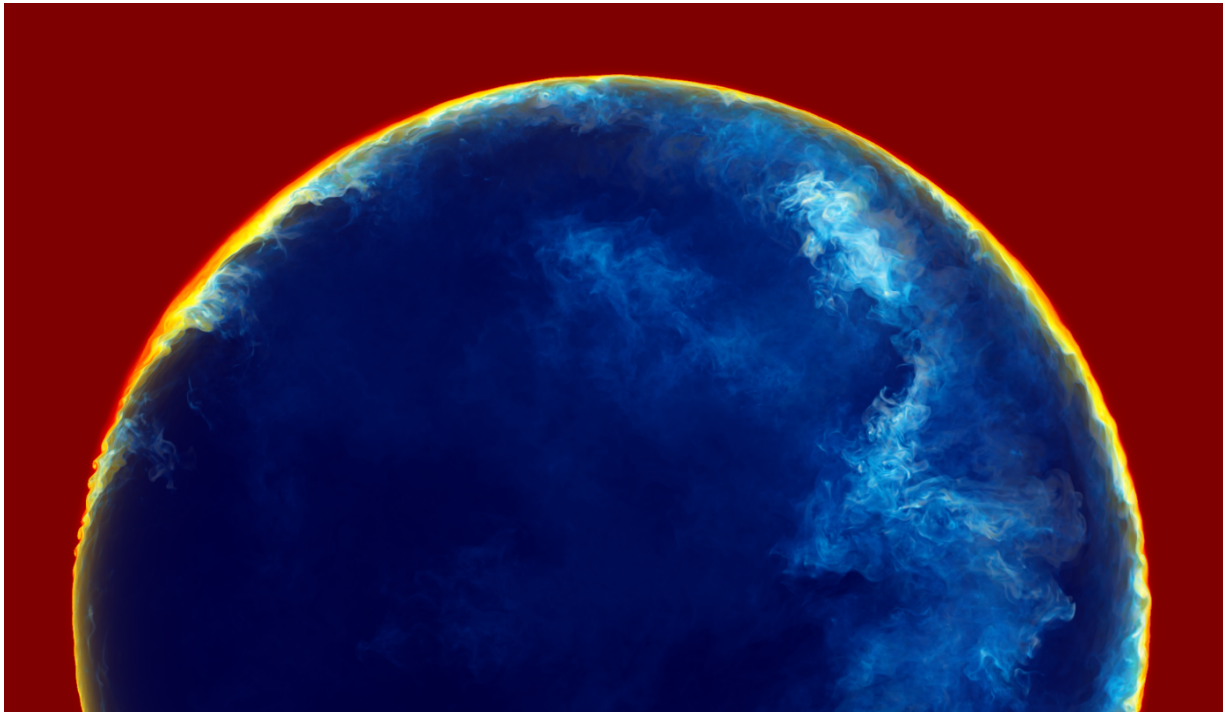

*Fig. 1. A zoomed-in view of entrainment of stably stratified, relatively buoyant gas from above the top of the convection zone at the center of a 25 M$_\odot$ model star that is rotating. The rotation rate at the convection zone boundary in this slice through the equator is roughly equal to the average speed of turbulent motions in the convection zone. The log of the entrained fluid concentration is rendered here, with concentrations decreasing through red, yellow, white, aqua, blue and dark blue. This snapshot 26.2 days (for the star) into the simulation reveals a circulation flow global in scale, with the downward flow driven by the collision at the convection zone boundary of oppositely circulating large convective eddies in a roughly dipole flow pattern.*

These flows deep in stellar interiors pose several significant challenges to accurate 3-D simulation. First, the convection itself is a tiny effect, but one that has enormous consequences by moving material radially within the star. The Mach numbers range around 0.035, and in some cases can be quite a bit lower still. For the earth's atmosphere, this Mach number corresponds to a wind of about 25 mph – no hurricane, not even a squall, but enough to move things around. The entrainment of gas at the convection zone boundary is an even tinier effect. An entrained concentration as low as 1 part in 100,000 can have a significant effect, while a concentration of 0.001 can have consequences that are simply enormous. A reason for this surprising importance of tiny concentrations is the nonlinear positive feedback arising from entrainment and burning of the ingested fuel. All these unusual features of these stellar hydrodynamic problems require special design of the numerical algorithm.

The low Mach numbers demand high-order interpolations of cell interface values of variables in our explicit PPM gas dynamics scheme (see [8]). At low Mach numbers, the accuracy of the advected amounts of different quantities is dominated by the accuracy of the interpolation of the cell interface value. Our PPM scheme estimates these critical values with one order higher accuracy than it delivers for the interpolation parabola as a whole. It is also true that in this low Mach number regime, if we apply monotonicity constraints upon the interpolation parabola when they are not needed, we can destroy the benefits of the special effort we make to evaluate the cell interface values with such care.

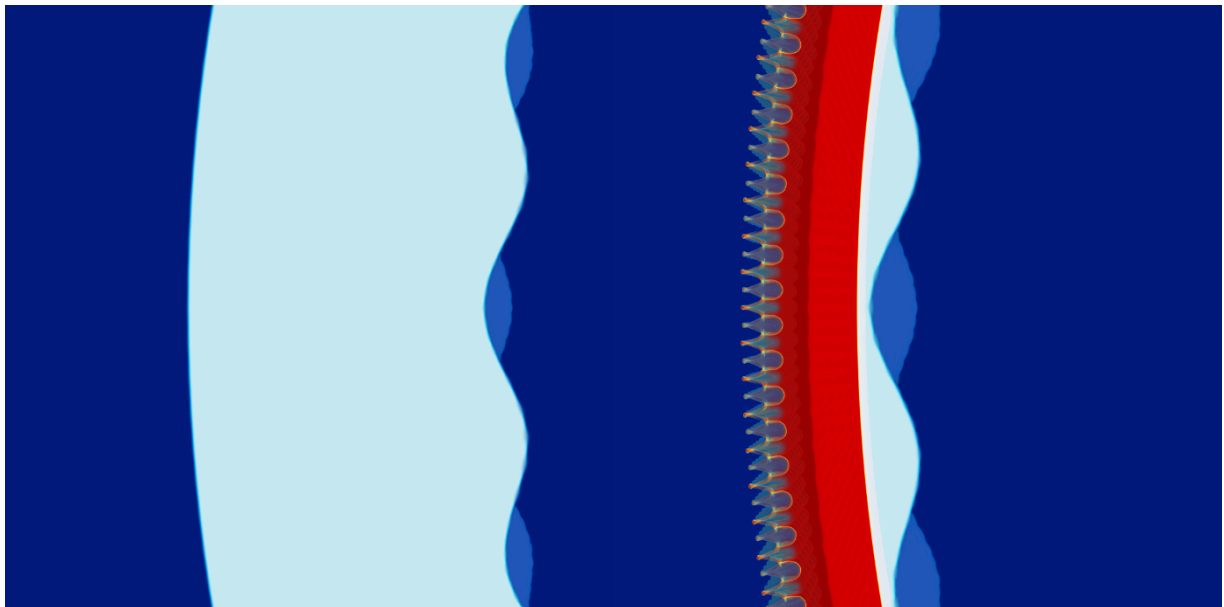

*Fig. 2. Zoomed-in views, initially (left) and subsequently (right), of the dense gaseous shell from the ICF test problem study discussed in the text (see [9]). Here we inspect the small portion of the shell in which we would anticipate the most troublesome Cartesian mesh imprint phenomena to appear. No such imperfections are visible, because each surface was disturbed initially by a very high frequency and very low amplitude perturbation that only comes into view in the later image, where it has been dramatically amplified by the Rayleigh-Taylor instability of the outer surface of the shell. These images show density, and the red region has been compressed by a shock moving to the right in the image. The grid in this simulation had over a trillion cells, but the benefits of this approach can be enjoyed on any size grid capable of accurately capturing the flow features of interest. See [9] for a more complete series of images from this study.*

As a result, we take the trouble to always evaluate the degree of smoothness of the function in a 5-cell portion of the grid centered on the cell of interest. When we judge the function to be smooth, we apply no monotonicity constraints. This procedure is well worth its cost, because it preserves the accuracy of the scheme as the Courant number approaches very small values. In flows where shocks can arise, we also introduce a fully 3-D dissipation inside shocks, which eliminates multiple grid imprint effects that we would otherwise observe in such flows (see [9]).

Aside from the above algorithmic accommodations, which we have made in all our codes since about 1990, we have made two other, major accommodations for our stellar hydrodynamics problems. The first is to save the initial, spherically symmetric state of the star on our grid, and then to subtract it out from the numerical scheme very carefully. We advance physical state variables that give perturbations to this initial state using nonlinear equations that are valid for perturbations of any size. Because the initial state is very rapidly varying in the radial direction, this lifts from the numerical method the burden of constantly evaluating the derivatives of this initial state with such high accuracy that small differences from those values can properly drive the dynamics of the convection flow. We take great care to explicitly cancel out all large contributions from the initial state to differences of values that are very nearly equal. This not only increases the accuracy of our computation, but it also allows that computation to be carried out with only 32-bit precision. This has an enormous benefit, not least because the speed of execution of the code is roughly doubled. This technique of subtracting out an unperturbed state was included in our earlier work on Rayleigh-Taylor instabilities [10], but the context of our present stellar hydrodynamics problems makes this subtraction much more important.

A second algorithmic accommodation we have made is to track the concentration of the entrained gas using the PPB moment-conserving advection scheme. We have described this scheme in detail elsewhere [11,8]. It is inspired by van Leer's Scheme VI from the 1970s [12], with multiple modifications that make it multi-D, robust, and efficient. We apply this scheme and our PPM gas dynamics on a uniform Cartesian grid. The advantage of this grid is that, for a general flow problem, it has geometrically optimal properties that produce benefits in the ease and accuracy of directional splitting techniques as well as in MPI messaging patterns and interconnect bandwidth demands. However, our stellar problems are generally spherical, which does not match the Cartesian geometry well. Nevertheless we argue that once perturbations to our spherically symmetric initial state develop and grow into the nonlinear regime, the geometrical properties of the grid become largely irrelevant. We have addressed questions on this issue through multiple demonstrations and code comparisons. One such example is the 2-D study reported in [13]. A 3-D study, based upon a similar inertial confinement fusion (ICF) test problem, was reported in [9]. In that latter study, we specially constructed a problem involving multifluid interfaces and their instabilities in the context of a perturbed spherical implosion for which grid imprint effects arising from our Cartesian mesh would become immediately visible at a glance. We found that as long as the unstable multifluid interfaces in that ICF problem were given high frequency initial perturbations that were visibly undetectable, no grid imprints of any significance developed. In that study, we were able to increase the grid resolution to $10560^3$ cells, so that convergence under grid resolution was well established. This approach to inhibiting grid imprint effects in our numerical treatment is illustrated in Fig. 2, from our study in [9]. This can be compared to the zoomed-in view of the region near the top of the convection zone in one of our recent stellar hydrodynamics simulations that is shown in Fig. 1.

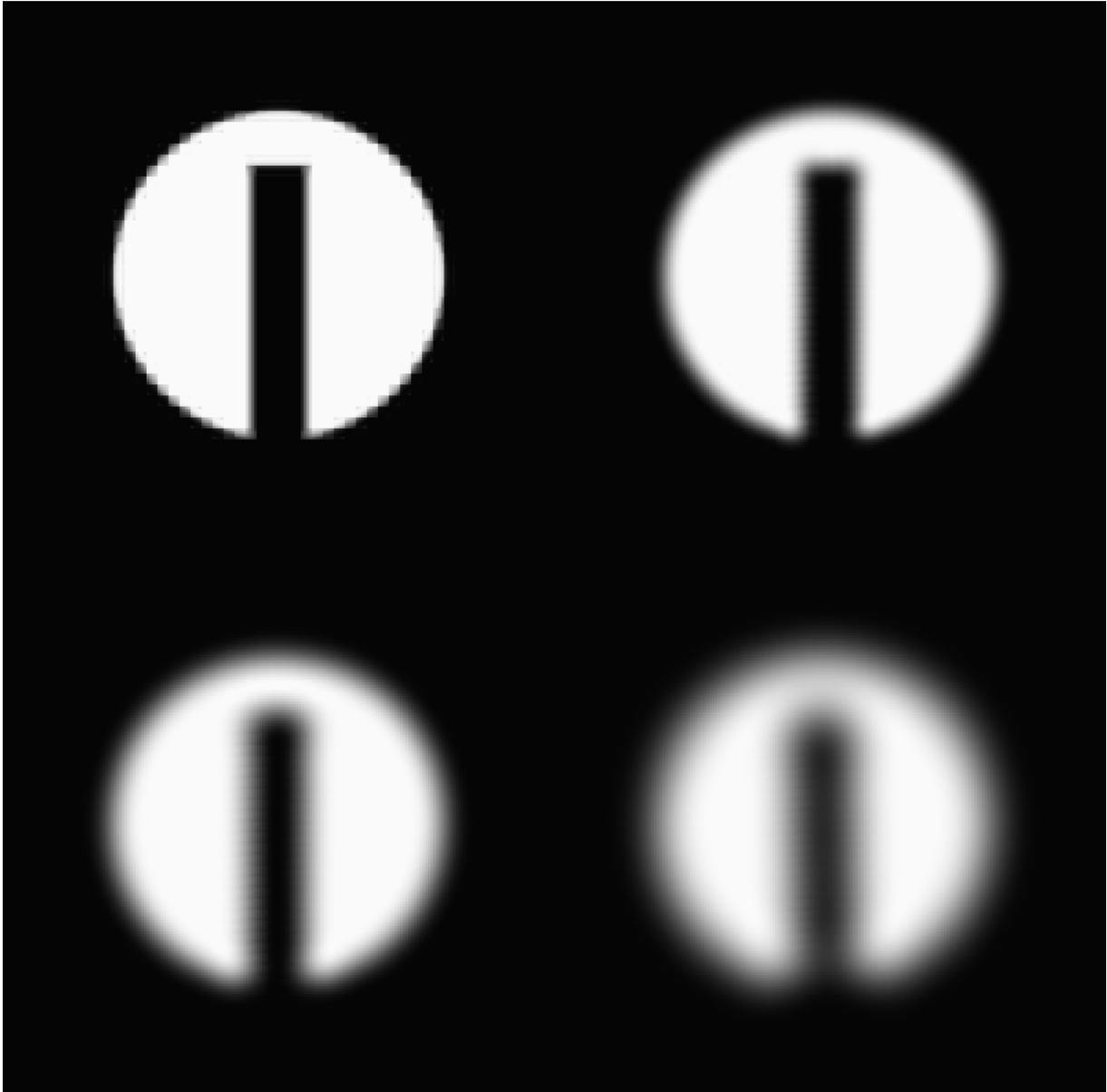

*Fig. 3. Results of the standard advection problem originally devised by Zalesak. We fill a slotted circle with gas whose fractional volume in grid cells we track with our PPB advection scheme in 2D on a uniform Cartesian grid of 100×100 cells. Only the upper central 50×50 cell portion of this grid is shown. A velocity field is applied which causes the gas to rotate around the center of the entire grid as a solid body. The fractional volume of tracked gas is shown at 0, 1, 10, and 100 revolutions of the rotational flow. Each grid cell is represented by a 4×4 square of subcells in which the fractional volume is computed using the 6 low-order moments of this variable that are advanced in a conservative manner by the PPB scheme. In the initial state, at the top left, the fractional volumes and higher moments are computed exactly for a shape with boundaries of zero thickness. Grid imprints are clearly evident. By one revolution, these have vanished.*

To drive home the relevant numerical benefits of our PPB multifluid advection, we show results in Fig. 3 of a standard advection test problem. This test problem, which follows a multifluid boundary embedded in a rotational flow through many rotations, is quite relevant to our convection flows in stars. The main difference, however, is that this is linear advection, in which the length of the multifluid interface does not increase. It is a fundamental property of unstable interfaces that their lengths (surface areas in 3-D flows) increase. That causes a stretching of the interfaces, and this acts as a stabilizing mechanism that tends to keep our numerical representations of such interfaces as sharp as the numerical scheme allows it ever to be. This effect can be clearly seen in our study of Rayleigh-Taylor instability in [10] and [14] and is a benefit visible in action in Figs. 1 and 2. It is a feature of PPB advection that constraining the representation of the multifluid volume fraction in a cell is almost never performed, and this makes the numerical representation relatively free from glitches that can set off secondary interface instabilities, as can be seen under unfavorable circumstances with the much simpler but still highly accurate PPM advection technique.

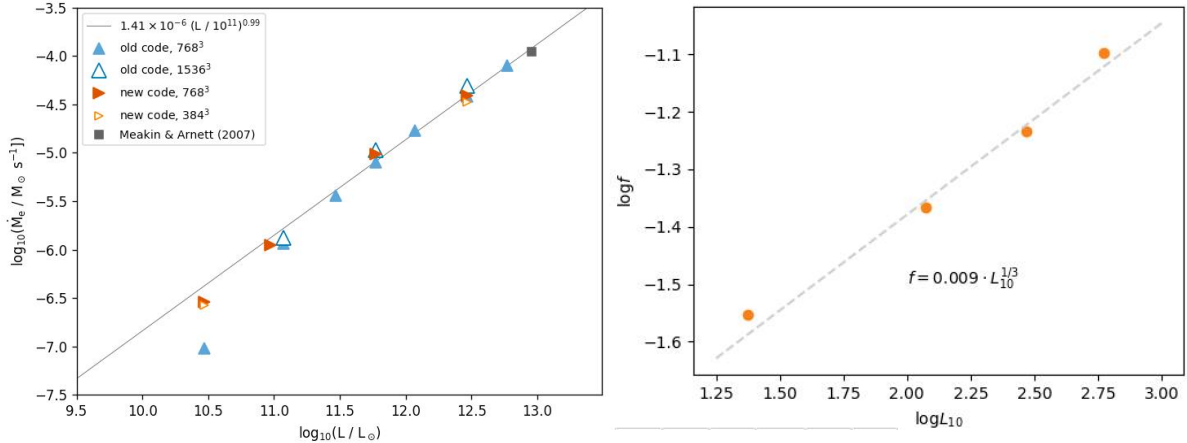

*Fig. 4. Scaling laws for entrainment rate (left) and mixing region thickness (right) as determined by 3-D PPMstar simulations of convective boundary mixing at the top of the convection zone above the O-burning shell in model 25 $M_\odot$ star. Note in the left-hand plot that our new code converges more rapidly than the old one as the grid is refined and gives results at low luminosities that are more in keeping with higher luminosity runs. This, we believe, is due to our subtracting out the unperturbed state of the star.*

## 2. Core convection in rotating main sequence stars.

We have reported work on stellar hydrodynamic problems involving ingestion of nuclear fuels at convection zone boundaries in multiple articles since 2012 [3,8,15-18,36]. This work can be compared to multiple efforts of other investigators using quite different numerical approaches [19-23,37,38]. We believe the principal advantages of our approach described above come from our ability to embody these techniques in a code that scales to extreme levels and also runs very well on each of the thousands of multi-CPU nodes it uses. On the Blue Waters machine at NCSA, and now also on the Niagara machine on Compute-Canada's SciNet, we are therefore able to simulate the entire central portion of a model star for a run of a million time steps or more in just about a day, running at around 20 time steps per second. We use modest grids of only $1536^3$ cells, and standard running configurations of 7344 nodes on Blue Waters and 1088 nodes on Niagara. The algorithm is explicit, and therefore we prefer to run problems for which the maximum Mach number encountered on the grid is at least about 0.035.

In this section, we present preliminary results of simulations on Blue Waters of core convection in massive main sequence stars. In this case, the Mach numbers are very low, but we address this issue by increasing the driving luminosity by a factor of 1000, which causes the velocities in the flow to increase by a factor of 10, bringing them into a cost-effective range for our code. We have established these scaling relationships in multiple series of simulations for different flows of this nature (see, for example, [24] and [17]). The scaling relationships can be understood by the following plausibility arguments. The entrainment rate scaling linearly with driving luminosity suggests that it is not the details of the flow near the convective boundary that determine the entrainment rate, but instead the energy available in the convection flow that can be used to drag down relatively buoyant gas. With more energy, we get more fluid dragged down. The argument for the scaling of the convective velocity cubed with the driving luminosity is as follows. The luminosity determines the size of the kinetic energy flux in the convection zone. Our extensive series of simulations of convection in the 1990s [25,26] indicate that this flux is a roughly constant fraction of the net convective energy flux. Increasing the overall energy flux in the layer by a factor of 10 would increase the kinetic energy flux by this same factor. The convection zone is fully turbulent in all our star models. Consequently, in the bulk of this zone, all 3 velocity components have, on average, the same rms value. The kinetic energy flux is proportional to the product of 3 such rms values, and therefore each of these 3 scales with the 1/3 power of the driving luminosity. The scaling relationships are observed results of our studies. We can use the above arguments to understand them, even though these arguments are not rigorous. Evidence for these relationships is shown in Figs. 4 and 5.

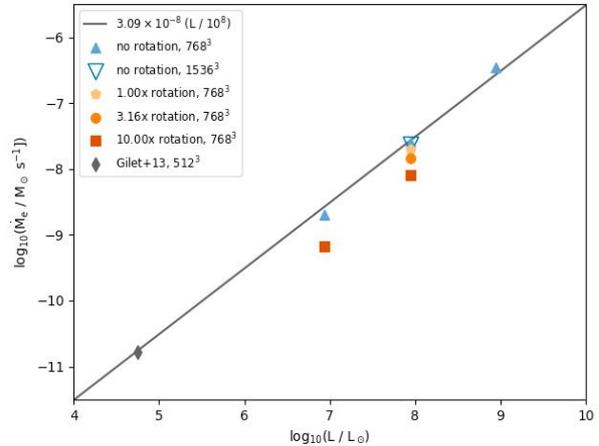

*Fig. 5. Scaling of various core convection runs with luminosity. Note that the simulation of Gilet et al. [22] lies on the scaling trend, despite its separation by 2 orders of magnitude from our runs and also its use of a very different, anelastic technique.*

## 3. A 25 $M_\odot$ rotating stellar model.

Exploiting the scaling relationships reported above, we consider core convection in a 25 $M_\odot$ rotating star. We treat the rotation by adding a solid body rotational flow component to our simulation gradually over the first several thousand time steps of the run. We stop this process when we reach a desired rate of rotation at the top of the convection zone that is centered on the core of the star. Our boundary condition is that our star is enclosed in an impenetrable and immovable sphere that is inscribed in the cubical domain of our uniform Cartesian grid. We make no special effort to have this confining sphere be frictionless, since in our problems there is usually very little material, relatively speaking, located in the region of this sphere. As an additional simplifying device, we let the gravity holding the star together drop to zero a few cells before the confining sphere is reached. This allows us to use a very simple boundary condition at the sphere. The extent to which any aspect of this boundary treatment matters in the simulation should be evident from the simulation results. It is very important to note that our confining sphere is *not* holding the star together by pushing on it radially with a pressure. Instead, our star is held together by gravity, and the pressure at the location of our confining sphere is essentially negligible in most of our problems.

The main interest in studying massive main sequence stars in this way is to obtain accurate main sequence lifetimes for such stars. These stars can be observed via asteroseismology using recent satellites launched to locate exoplanets. Ingestion of hydrogen fuel from the top of the convection zone can keep the core hydrogen fire burning longer than would otherwise be possible. When the star is not rotating, we find a global convection flow that is dominated by the $l = 1$ spherical harmonic, with a dipole convection pattern. This is the same sort of flow that we found 2 decades ago in simulating the deep convective envelope of a giant star [26]. That earlier study explored the effect of the spherical geometry on convection. In the present work, we have such deep spherical convection flows interacting with rotation. This is a situation that has been studied previously by Toomre and his collaborators [27-29], with a special focus on the case of the sun, where the convection zone does not extend downward all the way to the center of the star. They take an anelastic approach in a spectral code that includes magnetic fields, which are of great interest in the case of the sun. Here we have a simple treatment, with our uniform Cartesian grid and zero magnetic field.

The snap shot of the flow in Fig. 1 is taken from a simulation in which the rotation speed at the top of the core convection zone is roughly equal to the rms average convective velocity in the non-rotating case. Here the flow is modified by the rotation, but not so very much. We show results from cases rotating 3.16 times and 10 times faster in Fig. 6. These model stars rotate fast enough for the Coriolis forces to be significantly felt (although we perform the simulation in an inertial frame, where these "fictitious" forces do not apply). We can see in Fig. 5 that the entrainment of gas from above the convection zone is significantly reduced for rotating stellar models, yet the entrainment is still quite effective. Just as the convection tends to produce a well-mixed region of the star, it also transports and mixes the specific angular momentum. Although both simulations shown in Fig. 6 have been carried through multiple turn-over times for their large gyres, longer runs, ideally also carried out in rotating coordinate systems, will be necessary to study the redistribution of angular momentum by the convection in detail. Such runs are planned in the future. They should be made far more cost effective by our implementation of our code on GPU-accelerated nodes of large machines like Summit at the Oak Ridge National Laboratory. How we plan to do this is discussed in the next section.

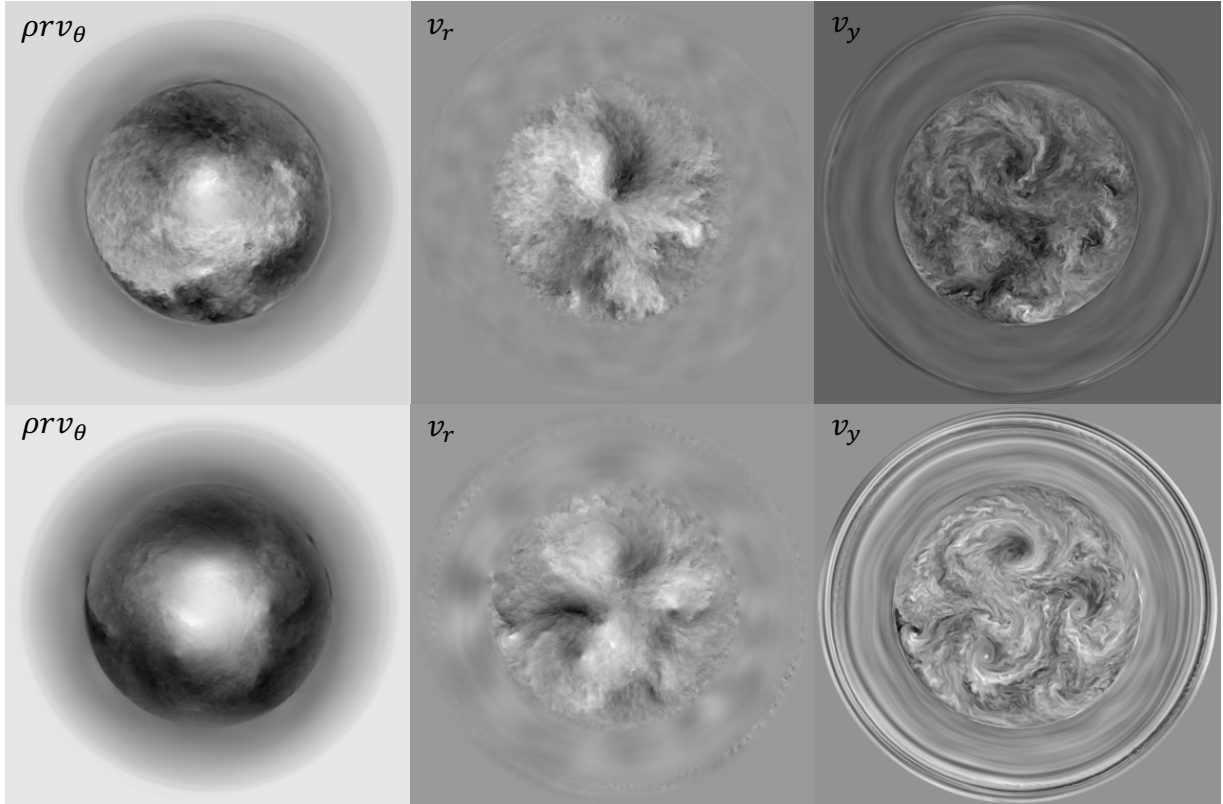

*Fig. 6. Snapshots from simulations of the same 25 $M_\odot$ main sequence stellar model that is shown in Figure 1, but on the top row here the star rotates at 3.16 times the rms velocity in the convection zone and on the bottom row it rotates at 10 times that velocity. These rotation rates apply at the top of the core convection region, a boundary that is clearly seen in all these images. Further out in radius, the location of our bounding sphere, which is not rotating, is visible. The images at the left show the angular momentum in the equatorial plane. The star is rotating clockwise, so that its highest rotational angular momentum is black, while smaller values are shades of gray to white. The central images show the radial component of the velocity in the equatorial plane, with white showing outward motion and black inward. Large gyres are evident that span the convection zone between the center of the star and the convection zone top. The images at the right show the component of velocity out of the page. Here the large gyres are particularly easily made out. These have the appearance of hurricanes, and their spinning is caused by the same mechanism.*

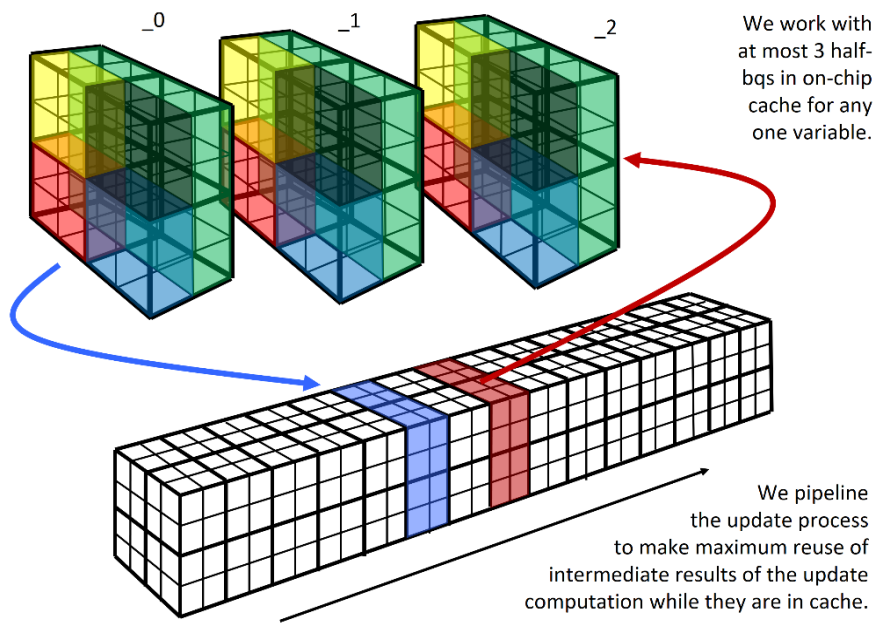

*Fig. 7. Diagram of the grid pencil update process. Each half-briquette record consists of 4 mini-briquette records, as explained in the text. Execution proceeds in the direction of the arrow, with 3 half-briquettes in the process of being updated at any one time using only cached data. Each half-briquette record is 2 KB.*

### 4. Implementing the PPM+PPB algorithm for efficient execution on GPUs.

The PPM gas dynamics algorithm [30,11] in the form which we now use [31] has been augmented with PPB [11,8] advection of a fractional volume variable, `fv`, that gives the fraction of the volume of a cell that is occupied by a special, tracked multifluid constituent. The cell average of this variable and its 9 lowest-order moments are updated by the algorithm in a directionally split technique (see [8]). Here we will not repeat the details of these algorithms. We simply note that we compose all the steps necessary to update grid cells for a single 1-D pass into a single, massive, vectorized loop that extends for thousands of lines of Fortran code. For decades, we composed such algorithmic steps as a series of subroutines, each operating upon either single strips of grid cells or bundles of such strips, which we called *grid pencils*. In such a classic implementation, vector lengths were equal to the number of grid cells in a strip, plus a few extra cells on each end. Each subroutine consisted of a series of vector loops with progressively fewer and fewer of these extra cells contained in its vectors. Because this is a difference scheme, these vectors could not be aligned. For processing on devices manufactured after about 2000, these loops had to be "strip mined" by the compiler/hardware into a series of loops with much shorter vectors, and those vectors had each, in general, to be shifted to produce the perfect alignment required by SIMD engines, the main computational engines in all devices manufactured today. This code structure produced long vector loops containing the necessary parallelism. However, the processor cache rapidly fills up with the long vectors and vector temporaries needed to implement such loops running at a reasonable speed in the absence of sufficient memory bandwidth. Consequently, only a few of these loops can be executed out of cached data, and there must therefore be a great deal of data traveling unproductively back and forth between the processing chip and its directly connected off-chip memory in order to perform all the many stages of the numerical algorithm.

## Performance gains

### Redundancy in computation eliminated

| | Workspace / thread (KB) | flop/cell | | |
|---|---|---|---|---|
| | | Fortran-W | Pipelined | % redundancy |
| RK-adv | 16.59 | 379.89 | 162.92 | 133.16 |
| PPM-adv | 19.2 | 454.61 | 273.31 | 66.34 |
| tp3 | 208.28 | 5195.77 | 3218.67 | 61.43 |

### Performance gains for PPM-adv

| | Speed-up from | | |
|---|---|---|---|
| | briquettes | pipelining-for-reuse & memory reduction | both |
| Nehalem | 2x | 3.33x | 6.69x |
| Sandy Bridge | 3.78x | 1.66x | 6.28x |

Nehalem : Xeon 5570 ; Intel 9 Fortran Compiler; 16 OpenMP threads running on two sockets
Dual-socket, 4-core @ 2.93GHz, SSE-4.2 (128-bit)
Sandy Bridge : Xeon ES-2670; Intel 13 Fortran Compiler; 32 OpenMP threads running on two sockets
Dual-socket, 8-core @ 2.6GHz, AVX (256-bit)

*Table 1. Results of a study of performance in 2014 by Jagan Jararaj working with our team (cf. [34]) that quantifies performance increases resulting from the 2 principal code restructuring techniques discussed here for our PPM codes. The first performance gain comes from the short, aligned vector operands provided by the briquette data structure, and the second comes from pipelining of the briquette update to remove redundant computation and reduce the size of the on-chip data workspace. No changes in recent hardware act to reduce this overall benefit of over a factor of 6. GPUs of the latest design can also benefit from these code structuring techniques, because we have reduced the on-chip data workspace requirement from 208 KB in this table to meet their limit of 32 KB plus an additional 16 KB of "shared memory" in Nvidia's Volta GPU.*

Our present code organization has resulted from our work with the Cell processor nodes of the Los Alamos Roadrunner machine [32-34,10]. We retain the concept of the grid pencil, but, as shown in Fig. 7, we update it progressively, one pair of 2 cross-sectional grid planes of 4×4 cells at a time. To facilitate this process, we store the physical state variables for each 2×2×2 cell mini-briquette together as a contiguous mbq-record in main memory. For an *x*-pass, with *i*, *j*, and *k* denoting *x*, *y*, and *z* indices, the mbq-record would be represented as a Fortran array indexed as mbqrec(j,k,i,iv), with the iv-index ranging over the 16 physical state variables. Such records are composed in 2×2×2 arrangements into briquette records, or bq-records, in memory so that they would be indexed in Fortran as bqrec(j,k,i,iv,jmbq,kmbq,imbq), a 7-dimensional array. Each of the 2 values of imbq, the *x* mbq-record index, corresponds to a different half-briquette, the primary data entity for our processing. The composition of each bq-record in main memory out of 8 mbq-records allows us to transpose this data for processing in the *y*- or *z*-passes by transposing the contents of each mbq-record, and then transposing the arrangement of these records in the bq-record. All these transpositions can be performed while the relevant data is on the processing chip, so that the cost is small, although the code involved is significant and, unfortunately, confusing. The positions of the bq-records themselves inside the grid brick data structure remain unchanged in the transposition process.

We go to a great deal of trouble in packing our data into records so that it can be easily and efficiently prefetched, as indicated in Fig. 7, long before its presence on the chip is required. Once we demand that it must have landed, we must rearrange the contents to form 32-word vectors of individual physical state variables. We do this so that the contents of each vector line up as shown in Fig. 7, with the effective Fortran structure as in var(j,k,jmbq,kmbq,i), although of course we almost never expose all these interior dimensions of the data. We have revised our PPM+PPB algorithm so that the difference stencil extends outward only 4 cells in each direction along the *x*-axis for the *x*-pass. This means that we never need to hold in the on-chip cache more than 3 of our 32-word vectors for any quantity. With GPUs in mind, which have forced us to use 32-word rather than 16-word vectors, we do not add a final imbq-dimension. Instead, we manage separate vectors to which we append suffixes _0, _1, _2 as shown in Fig. 7. Each time we prepare to process a new half-briquette, we need to overwrite many of our vectors to accomplish a barrel shift of the data. This sounds like more trouble than it entails, and as far as our testing indicates, its cost is small. The speed of our implementation on no device now available is significantly limited by these data manipulations. Instead, it appears to be limited by the ability of the devices to perform our arithmetic at a good clip when all necessary operands are in the on-chip data store. Our codes run very well, as one can see from Tables 1 and 2, but of course they could run better. To our knowledge, our code's performance is not now limited on any device by insurmountable obstacles such as the speed of light, memory bandwidth, or the size of the on-chip cache.

If we consider that at the outset of a half-briquette update we have all our 32-word vectors prepared and ready to go, then the code to perform that update should be a long, uninterrupted set of 32-wide SIMD arithmetical instructions, where all operands are immediately available on the chip. Our algorithm performs between about 1300 and 1400 such 32-wide SIMD flops to update a half-briquette, so this loop should not be outrageously long, in principle. However, for increased code readability and for convenient management of the large group of vector temporaries involved, we use subroutines that call other subroutines, so that, overall, this code module increases to 6 or 7 thousand Fortran lines. About half of this code implements computations that we do only when a data output dump will follow this particular pass. The management of temporary vectors via subroutine stacks that disappear upon exit is the largest inconvenience in this programming style. It is this technique that has enabled us to bring the size of our on-chip data workspace down from the 208 KB noted in Table 1, from 2014, to 32 KB of a GPU "register file" plus 16 KB of GPU on-chip "shared memory." This aggressive data workspace reduction is unnecessary for CPUs, which have sufficient cache storage space, but is essential for GPUs, which do not. In addition to straightforward arithmetic, we have groups of lines in which we must form new 32-word vectors consisting of the last 16 words of one vector and the first 16 of another. This is easy to express in Fortran. For CPUs, many of which process no more than 16 words at a time, these statements are unnecessary and should be removed by the compiler. For GPUs, these operations are expensive due, we think, to a special hardware feature, and they seem to benefit from doing 4 of them together, when possible. Writing a program in this style is unusual, but one can become accustomed to it, so that it is no more difficult than writing a program any other way. We have found that writing our codes this way enhances their performance on all devices we have seen.

**5. Code performance experience and projections.**
Because we have been running our codes almost exclusively on the Blue Waters machine at NCSA, which has only a small fraction of GPU-accelerated nodes, our experience with running our codes at scale is mainly with dual-CPU nodes. We have, however, packaged a testing version of the PPM+PPB package, written in the style described above, that runs on either a dual CPU node or on a GPU node in a performance testing mode. The performance it achieves in this mode is born out on CPU nodes in our experience when the code is scaled up to thousands of nodes running a single problem. The GPU test does not involve any data movement back and forth between the GPU's attached memory and that of the CPU host, but in a full code implementation, we do not expect these data transfers to slow us down much, because they will only contain MPI messages, which are small relative to the overall data being updated on the GPU. Our GPU implementation was described in detail in [35]. Below, we include performance information from that study, updated to include results for more recent hardware. We have retained the numbers for the older processors, because these are found in some large systems still in service today. An evolution over 6 years is shown in the table. Improvements over this time interval of factors of 6.8 for CPUs and 8.5 for GPUs are not too far below Moore's law expectations.

As we go to higher and higher performing devices in Table 2, it becomes progressively more questionable whether the interconnect on a large system will be able to keep up. This concern is especially acute with the latest generation of GPUs, which, according to Table 2, offer roughly double the performance of dual-CPU nodes. We can always decrease the demands on the interconnect by doubling the linear dimensions of the grid bricks that we update as uninterruptible tasks for MPI ranks in our codes. Each such doubling increases the computational work between communication events by a factor of 8, while the amount of data then communicated increases by only a factor of 4. Reducing demands on the interconnect bandwidth in this way can therefore increase the wall clock time for a run very significantly. However, if job queue wait times are long for requests involving large fractions of the machine, running the same problem on 8 times fewer nodes can turn out to be a good choice from the standpoint of higher efficiency of utilization of the requested resources and shorter wait times for the run to begin. On systems where there is higher effective interconnect bandwidth on special subsets of nodes, scaling a problem to this well-connected subset can be a good idea.

Table 2. multifluid PPM+PPB 1-D Grid Pencil Update

| Device | GHz | Threads of control / cores | Gflop/s (32-bit) | Required Continuous GB/sec |
|---|---|---|---|---|
| dual AMD Interlagos node | 2.3 | 32 / 32 | 71.7 | 8.30 |
| dual Intel Sandy Bridge | 2.0 | 32 / 16 | 84.7 | 9.80 |
| dual Intel Haswell node | 2.3 | 64 / 32 | 267 | 30.8 |
| **dual Intel Skylake node** | **2.4** | **80 / 40** | **415** | **47.9** |
| Nvidia K20 | 0.73 | 168 / 14 | 121 | 38.9 |
| Nvidia K40m | 0.875 | 120 / 15 | 140 | 15.7 |
| 1 Nvidia K80 | 0.875 | 208 / 13 | 248 | 27.8 |
| Intel Knights Landing node | 1.4 | 272 / 68 | 260 | 30.0 |
| Nvidia Titan-X Pascal GP102 | 1.800 | 448 / 56 | 610 | 68.4 |
| **Nvidia Volta GPU** | **1.800** | **640 / 80** | **1033** | **115.8** |

Our code is designed to give MPI messages a full grid brick update time interval to arrive at their destinations, as is described in some detail in [18]. The grid brick update consists entirely of a sequence of grid pencil updates, which are all independent in each of the three 1-D passes. We write the code so that we may begin the next 1-D pass before the present one is completed, and the dependencies between one pass and the next are minimized by the order in which the grid pencil tasks are launched. As a practical matter, we may therefore consider each grid brick update to consist of a large number of grid pencil updates that are, pretty much, independent. For a grid brick of $M^3$ grid briquettes, the cost of the update is $(M+2)^2 \times (M+1) + (M+2) \times (M+1) \times M + (M+1) \times M^2 = (M+1) \times ((M+2) \times 2 \times (M+1) + M^2)$ single grid briquette updates. We have found that, as a rule of thumb, one obtains excellent performance and efficiency if one chooses M equal to the number of CPU cores on the node. From Table 2, we see that for today's hardware, a choice of M=40 or M=80 is indicated. In these two cases, the grid brick update is 206,804 or 1,594,404 briquette updates of about 83,200 flops each. This amounts to 17.2 or 132.65 Gflops, and 0.0414 or 0.1284 sec. at the computation rates given in Table 2. In this time interval, $(M+2)^3 - M^3$ briquette records must both leave and come onto the node performing the grid brick update. With each briquette record being 4 KB, this is $2 \times 40.35$ MB or $2 \times 157.5$ MB. This requires a continuous, actually delivered link bandwidth to the node of 1.95 or 2.45 GB/sec. These are not impossible numbers, but they are pretty much at the margin of what is possible today. Note that we have included no consideration for finite latency of message passing. These estimates imply that although in principle better device design could deliver more performance to our codes at each node, there is no point in making such design improvements unless the interconnect bandwidth in today's machines is increased. There is a further consequence. If these grid brick sizes are close to the smallest ones that can be efficiently updated on today's equipment, with a minimum of two grid bricks assigned to each node, and if we demand that our problems scale to, say, 4000 nodes, this implies minimal grids of about $(20 \times M)^3$, or either $3200^3$ or $6400^3$ cells. By today's standards, these are very large grids.

The scientific problems that our team is addressing have no difficulty in making good use of this new capability to attack much larger and more complex scientific problems. As a new generation of computing systems is being put into place, we will continue to adapt our code design to exploit these systems to simulate 3-D phenomena that play crucial roles in stellar evolution during relatively brief periods, such as hydrogen ingestion flashes and mergers of nuclear burning shells.

**6. Acknowledgements.**

Simulations and code development discussed in this article were performed on the Blue Waters machine at NCSA under support from NSF PRAC awards 1440025, 1515792, and 1713200, as well as a subcontract from the Blue Waters project at NCSA. Partial support for the code development also came from contracts from the Los Alamos and Sandia National Laboratories. The scientific research described here was supported by the NSF through CDS&E grants AST-1413548 and AST-1814181 and through subcontracts from NSF's JINA-CEE center (PHY-1430152). Simulations were also carried out on Compute Canada's Niagara computer. FH acknowledges supported by an NSERC Discovery grant.